\documentclass[11pt]{article} 
\usepackage{graphicx}
\usepackage{dcolumn}
\usepackage{bm}
\usepackage{color}
\begin{document}

\title{Ensemble averaged Madelung energies of finite volumes and
  surfaces}
\author{Peter Kr\"uger\\
Graduate School of Engineering and\\
Molecular Chirality Research Center,\\
    Chiba University, Chiba 263-8522, Japan\\
pkruger@chiba-u.jp}

\date{\today}

\maketitle

\begin{abstract}
  Exact expressions for ensemble averaged Madelung energies
  of finite volumes are derived.
  The extrapolation to the thermodynamic limit converges
  unconditionally and can be used 
  as a parameter-free real-space summation method
  of Madelung constants.
  In the large volume limit,
  the surface term of the ensemble averaged Madelung energy
  has a universal form, independent of the crystal structure.
  The scaling of the Madelung energy with system
  size provides a simple explanation for the structural phase
  transition observed in cesium halide clusters.
\end{abstract}





\section{Introduction}
The cohesive energy of ionic crystals is dominated by
the electrostatic energy between ionic point charges, known as the
Madelung energy.
The calculation of Madelung energies is a mathematically non-trivial
problem because of the long-range nature of the Coulomb interaction.
The Madelung constant of basic crystal structures
was first successfully calculated by Ewald~\cite{ewald}.
In the Ewald method, the Coulomb interaction is divided into a short
range part for which the Madelung sum converges fast in real space
and a long range part which can be summed in reciprocal space.
The Ewald method is very accurate and widely used, but it
is numerically rather involved and relies on periodic boundary conditions.
Alternatively, the Madelung energy can be calculated through
direct summation in real space, which is numerically simpler and
can be used in finite system~\cite{baker} and non-periodic structures.
However, the lattice sums are only conditionally convergent, i.e. the
result depends on the summation order.
This reflects the physical fact that in a finite crystal,
the potential at an inner site can be changed at will by choosing particular
surface terminations~\cite{martin}.
In three dimensions, the Madelung sums diverge for the most natural,
shell-like summation order~\cite{borwein85}.
Divergence can be avoided in two ways.
In the first type of methods, the lattice is
divided into neutral cells of vanishing dipole moment~\cite{evjen}.
The sum over these cells converges absolutely 
because the quadrupole-quadrupole interaction decays as $1/r^5$.
However, dipole moment free cells generally involve fractional ions at the
corners and edges, and can be difficult to construct for
low symmetry systems~\cite{eugen}.
In the second type of method,
the summation is done in the natural order of increasing
distance but the system is neutralized at each step by adding a
background or surrounding sphere~\cite{wolf,harrison}.
An example is the Wolf method~\cite{wolf} 
which keeps only the short-range part of the Ewald method but
compensates each charge inside the summation sphere by an opposite
charge at the cut-off radius.

Here we consider ensemble averaged quantities in finite subvolumes
of a macroscopic system, especially the mean electrostatic potential
at the sites of a given ionic species.
This leads to an alternative definition of Madelung energies which
converges unconditionally as a function of system size for any
subvolume shape.
We derive an exact expression of the Madelung constants in
finite spheres.
The leading term in the expansion over inverse size is found to be
independent of the crystal structure.
As a consequence, the Madelung contribution to the surface energy,
averaged over surface orientations, is the same for all crystal structures,
and provides a universal first order approximation of the
surface energy of ionic systems.
  We apply the theory to the relative stability of CsCl and NaCl ionic
  structures as a function of system size. Under the assumption of
  equal nearest neighbor distance we find that the stable structure
  switches from NaCl to CsCl when the system size exceeds a few hundred ions,
  which explains the phase transition observed in cesium halide clusters.
\section{Average Madelung energy of finite volumes}
We consider a collection of $N$ point charges at positions
${\bf r}_{i}^{\alpha}$, where $\alpha$ labels the different
species with charge $q_\alpha$.
The electrostatic energy is given by
\begin{equation}\label{u0}
  U= \frac{1}{2} \sum_{j\beta\ne i\alpha}
  \frac{q_\alpha q_\beta}{|{\bf r}^\alpha_i-{\bf r}^\beta_j|}
  =\frac{1}{2} \sum_\alpha q_\alpha \sum_{i}\phi({\bf r}^\alpha_i)
\end{equation}
where 
\begin{equation}\label{phii}
  \phi({\bf r}^\alpha_i)= \sum_{j\beta(\ne i\alpha)}
  \frac{q_\beta}{|{\bf r}^\alpha_i-{\bf r}^\beta_j|} \;.
\end{equation}
is the electrostatic potential at site~${\bf r}^\alpha_i$.
For crystals we take each ion in the unit cell as
a different species~$\alpha$ and $i$ is the cell index.
In the infinite crystal, 
$\phi({\bf r}^\alpha_i)$ is independent of $i$.
The Madelung constant of the species $\alpha$ is commonly defined as
\begin{equation}\label{Mai}
  M_\alpha = -\phi({\bf r}^\alpha_i){d}/{q_\alpha} \;,
\end{equation}
where $d$ is the nearest neighbor distance.
In a practical real space summation, one computes the potential
$\phi({\bf r}^\alpha_i)$ at some site
$i$ of a finite cluster (usually the central site),
and lets the cluster size $N$ go to infinity.
The problem with this approach is that the series in Eq.~(\ref{phii}),
converges only conditionally for $N\rightarrow\infty$.
This means that the sum, i.e. the potential at site~$i$,
depends on the summation order and may diverge.

Instead of the potential at a given site~$i$, we 
consider the average potential at the sites of the $\alpha$-ions,
\begin{equation}\label{phi}
  {\bar \phi}_\alpha =\frac{1}{N_\alpha}
  \sum_{ij\beta}^{i\ne j} \frac{q_\beta}{|{\bf r}^\alpha_i-{\bf r}^\beta_j|}
\end{equation}
where $N_\alpha$ is the number of $\alpha$~ions in the cluster.
The electrostatic energy in Eq.~(\ref{u0}) can be rewritten as
\begin{equation}\label{u}
 U =\frac{1}{2} \sum_\alpha N_\alpha q_\alpha {\bar \phi}_\alpha \;.
\end{equation}
Equations~(\ref{u0}) and (\ref{u}) are equivalent and hold for finite
and infinite systems.
For an infinite crystal, we obviously have
${\bar \phi}_\alpha=\phi({\bf r_i^\alpha})$.
This suggests a redefinition of the Madelung constants as
\begin{equation}\label{Mav}
  M_\alpha =  -{\bar \phi}_\alpha{d}/{q_\alpha} \;.
\end{equation}
Of course, as with Eqs.~(\ref{phii},\ref{Mai}),
the true Madelung constants are 
obtained in the limit $N\rightarrow\infty$.
The advantage of definition~(\ref{Mav}) is that 
the electrostatic energy can be expressed
simply in terms of $M_\alpha$ as 
\begin{equation}\label{U}
U = -\frac{1}{2d}\sum_\alpha N_\alpha M_\alpha q_\alpha^2 \;,
\end{equation}
which holds for any system, finite or infinite.

\begin{figure}[htbp]
 \begin{center}
\includegraphics[width=1.2\columnwidth]{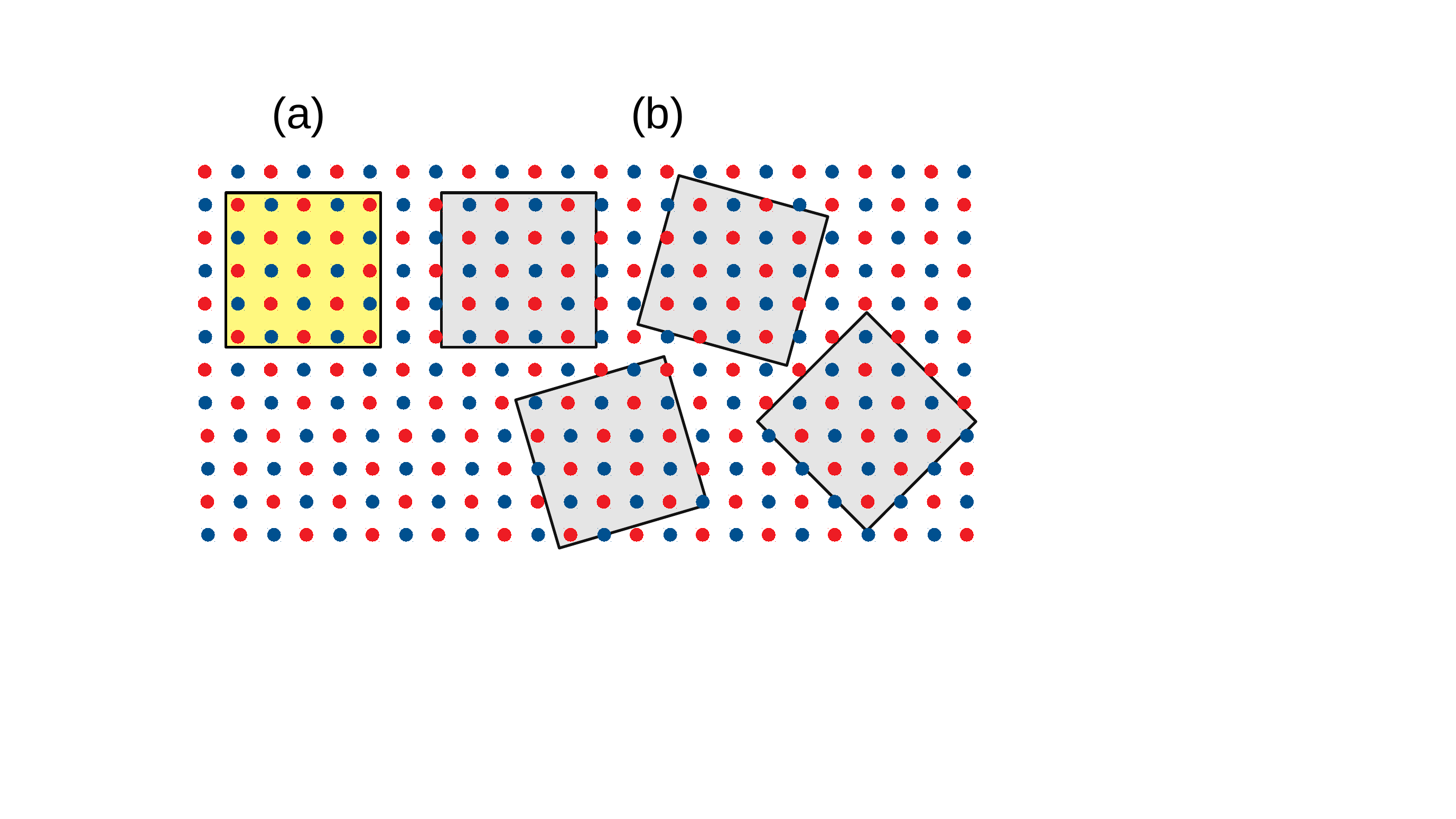}
\caption{
  Illustration of the difference between a standard real-space
  summation method (a), and the present, ensemble averaged method (b).
  Red and blue dots are positive and negative point charges and the
  black lines indicate the boundary of the finite cluster volume~$V$.
  In (a) the center and orientation of the volume~$V$ is fixed.
  In (b) an average is done over all possible
  positions and orientations of~$V$.
}\label{fig:scheme}
\end{center}
\end{figure}

We introduce the particle density of species~$\alpha$
\begin{equation}\label{rho}
\rho_\alpha({\bf r})=  \langle\sum_{i} \delta({\bf r}-{\bf r}_i^\alpha)\rangle
\end{equation}
where $\langle \dots \rangle$ denotes a statistical ensemble average.
The pair distribution function is defined as
\begin{equation}\label{g}
  g_{\alpha\beta}({\bf r},{\bf r}')
  =\frac{1}{\rho_\alpha({\bf r})\rho_\beta({\bf r}')}
   \langle\sum_{ij}^{i\ne j} \delta({\bf r}-{\bf r}_i^\alpha)
    \delta({\bf r}'-{\bf r}_j^\beta)\rangle  \;.
\end{equation}
With these functions, we can rewrite the ensemble average of the
Eq.~(\ref{phi}), as
\begin{equation}\label{phiav}
\langle{\bar \phi}_\alpha\rangle = \frac{1}{N_\alpha}
  \sum_{\beta} q_\beta \int d{\bf r}\rho_\alpha({\bf r})
  \int d{\bf r}'\rho_\beta({\bf r}') 
  \frac{g_{\alpha\beta}({\bf r},{\bf r}')}{|{\bf r}-{\bf r}'|} \;.
\end{equation}
Now we consider a finite subvolume~$V$ of an infinite ionic
solid and average over all possible positions
and orientations of~$V$.
Equivalently we can keep the subvolume
fixed in space and perform statistical averaging in an ensemble of crystals
of random position and orientation,
which is realized in a powder sample.
This ensemble is homogeneous and isotropic, i.e. it has the symmetry of
a fluid, where the density $\rho_\alpha$ is a constant, and
the pair distribution function $g_{\alpha\beta}(r)$
depends only on the distance
$r=|{\bf r}_i^\alpha-{\bf r}_j^\beta|$~\cite{mcquarry}.
As a result, and using $\rho_\alpha=N_\alpha/V$,
Eq.~(\ref{phiav}) simplifies to  
\begin{equation}\label{phi2}
\langle{\bar \phi}_\alpha\rangle 
=\frac{1}{V} \sum_{\beta}\rho_\beta q_\beta
\int_V d{\bf r}\int_V d{\bf r}'
\frac{g_{\alpha\beta}(|{\bf r}-{\bf r}'|)}{|{\bf r}-{\bf r}'|} \;.
\end{equation}
This expression holds for a finite subvolume~$V$ of any homogeneous
and isotropic ensemble, in particular for a fluid or powder sample.
Further, as explained above, statistical averaging in such an ensemble
is equivalent to considering a single crystal and
averaging over the position and orientation of the subvolume~$V$.
Hence Eq.~(\ref{phi2}) also holds in this sense.
In standard direct space lattice summation methods,
a finite cluster size corresponds to a subvolume of the infinite
crystal with fixed center and orientation
as illustrated in Fig.~\ref{fig:scheme}~(a).
In the present method, an average is done over
all possible positions and orientations (Fig.~\ref{fig:scheme}~b).
In the special case of a spherical subvolume considered in the
numerical applications
below, only positional averaging is meaningful.
However, the main results and conclusions are independent of the
subvolume shape.

For stability, any ionic system must
be {\em globally}\/ charge neutral, i.e.
\begin{equation}\label{qrho0}
  \sum_\alpha \rho_\alpha q_\alpha =0 \;.
\end{equation}
We do not assume any form of {\it local}\/ charge neutrality.
Averaging is done over all configurations in the ensemble,
including those where the finite volume~$V$ has a net charge.
By virtue of Eq.~(\ref{qrho0}), we may add to the integrand
in Eq.~(\ref{phi2}) any function $f({\bf r},{\bf r}')$
that is independent of~$\alpha$,
without changing the value of the expression.
We shall add $-1/|{\bf r}-{\bf r}'|$,
i.e. we replace $g_{\alpha\beta}(r)$ by the pair correlation function
$h_{\alpha\beta}(r)\equiv g_{\alpha\beta}(r)-1$
and obtain 
\begin{equation}\label{phi3}
  \langle{\bar \phi}_\alpha\rangle
  = \sum_{\beta} \rho_\beta q_\beta H_{\alpha\beta}^V\;,
\end{equation}
where
\begin{equation}\label{HV}
H_{\alpha\beta}^V =
\frac{1}{V}\int_V d{\bf r} \int_V d{\bf r}'
{h_{\alpha\beta}(|{\bf r}-{\bf r}'|)}{v(|{\bf r}-{\bf r}'|)} 
\end{equation}
with $v(r)=1/r$.
For $V\rightarrow\infty$, this simplifies to
\begin{equation}\label{Hinfty}
  H_{\alpha\beta}^\infty = \int_0^\infty h_{\alpha\beta}(r) v(r) 4\pi r^2 dr\;.
\end{equation}
While we focus on the bare Coulomb potential $v(r)=1/r$,
the present theory can in principle be applied to any potential form~$v(r)$.
For the special case $v=1$, $H_{\alpha\beta}^\infty$
are the Kirkwood-Buff integrals (KBI, usually denoted $G$ rather than $H$)
of solution theory~\cite{kb1952}.
Previously we have extended KBI theory to finite volumes~\cite{kruger13}.
We showed that the bad convergence of the usual, truncated integrals
can be avoided by an exact transformation from double volume integrals
to one-dimensional radial integrals.
Using this formalism~\cite{dawass17,kruger18}, we rewrite Eq.~(\ref{HV}) as
\begin{equation}\label{Hw}
  H_{\alpha\beta}^V   = \int_0^\infty h_{\alpha\beta}(r) v(r) w^V(r) dr
\end{equation}
where
\begin{equation}\label{wV}
  w^V(r)=
  \frac{1}{V}\int_V d{\bf r}_1 \int_V d{\bf r}_2
  \delta(r-|{\bf r}_1-{\bf r}_2|)
\end{equation}
is a weight function.
Note that Eq.~(\ref{Hw}) is actually a finite integral, since
$w^V(r)=0$ for $r>L_{\rm max}$ where $L_{\rm max}$ is the maximum
distance in~$V$.
Equations~(\ref{Mav},\ref{phi3},\ref{Hw},\ref{wV}) provide
an exact expression for the ensemble averaged Madelung potential
in finite volumes.
The weight function is conveniently expressed as
\begin{equation}\label{wVy}
  w^V(r)=4\pi r^2 y(r/L)
\end{equation}
where $L=6V/A$ and $A$ is the surface area.
The function $y(x)$ only depends on the shape of the volume~\cite{kruger18}.
For a sphere of diameter~$L$, the exact expression is~\cite{kruger13}
\begin{equation}\label{ysx}
  y_s(x) = (1-3x/2+x^3/2)\theta(1-x)
\end{equation}
where $\theta(x)$ is the unit step function
($\theta$=0 for $x$$<$0 and $\theta$=1 for $x$$>$0).
Analytic expressions of $y(r)$ are
also known for cube and cuboid~\cite{kruger18}. 
For any other shape, the function
can be easily computed numerically~\cite{dawass18}.
In order to accelerate convergence to the thermodynamic limit,
several extrapolations from the finite volume integrals
have been proposed~\cite{kruger13,kruger18,santos18}.
Here we focus on
the second order expression of Ref.~\cite{kruger18},
with weight function $u_2(r)=4\pi r^2 y_2(r/L)$, where
\begin{equation}\label{y2x}
  y_2(x) = (1 -23 x^3/8 + 3 x^4/4 + 9 x^5/8)\theta(1-x)
\end{equation}
For further reference, we also define
\begin{equation}\label{y0x}
  y_0(x) = \theta(1-x)
\end{equation}
which corresponds to simple truncation of radial integrals
or lattice sums at $r=L$.

\section{Application to crystals}
In the following, we consider a crystal with unit cell of volume $V_c$ 
containing $m$ point charges~$q_\alpha$,
$\alpha=1\dots m$, at positions ${\bf r}^\alpha$.
We take each atom in the unit cell as a different
species~$\alpha$, such that $\rho_\alpha=1/V_c$
for all~$\alpha$.
The spherically averaged pair distribution function,
appropriate for powder samples, is given by
\begin{equation}\label{gcrystal}
  g_{\alpha\beta}(r) 
  = \frac{V_c}{4\pi r^2}\sum'_{\bf T}
  \delta(r-|{\bf r}^\alpha-{\bf r}^\beta+{\bf T}|) \;,
\end{equation}
where ${\bf T}$ are lattice vectors and
the primed sum indicates that the term ${\bf T}=0$ is excluded
for $\alpha=\beta$.
Equivalently, Eq.~(\ref{gcrystal}) may be written as a sum over 
shells,
\begin{equation}\label{gshells}
g_{\alpha\beta}(r)
=\frac{V_c }{4\pi r^2} \sum_{k}
  n_k^{\alpha\beta}\delta(r-R_k^{\alpha\beta}) 
\end{equation}
where $n_k^{\alpha\beta}$ is the number of $\beta$ ions on
shell number~$k$ of radius
$R_k^{\alpha\beta}=|{\bf r}^\alpha-{\bf r}^\beta+{\bf T}|>0$
around an $\alpha$~ion.  
Inserting Eq.~(\ref{gshells}) into Eq.~(\ref{Hw}) we obtain
\begin{equation}\label{Habv}
  H_{\alpha\beta}^V =
  {V_c }\sum_{k} n_k^{\alpha\beta}v(R_k^{\alpha\beta}) 
  y(R_k^{\alpha\beta}/L)  - B^V
\end{equation}
where
\begin{equation}\label{bv}
  B^V = \int_0^\infty v(r) y(r/L) 4\pi r^2 dr \;.
\end{equation}
The term $B^V$ comes from the constant $1$ which was subtracted
when going from $g_{\alpha\beta}$ to $h_{\alpha\beta}$ in
Eqs~(\ref{phi2},\ref{phi3}).
By virtue of Eq.~(\ref{qrho0}) the $B^V$ terms cancel upon summing over~$\beta$
in Eq.~(\ref{phi3}). While $B^V$ has no direct physical meaning,
it is important for the convergence of the individual terms
$H_{\alpha\beta}$ in Eq.~(\ref{Habv}).
For a sphere of diameter $L$, we obtain
$B^V = 4\pi L^3 \int_0^1 v(xL) y_s(x)x^2 dx$ with $y_s$ given
in Eq.~(\ref{ysx}).
For the Coulomb interaction, $v(r)=1/r$, this yields
$B^V = 2\pi L^2/5=(12/5)V/L$. For $v=1$ (KBI) we have
$B^V=\pi L^3/6=V$.

\begin{figure}[htbp]
 \begin{center}
\includegraphics[width=\columnwidth]{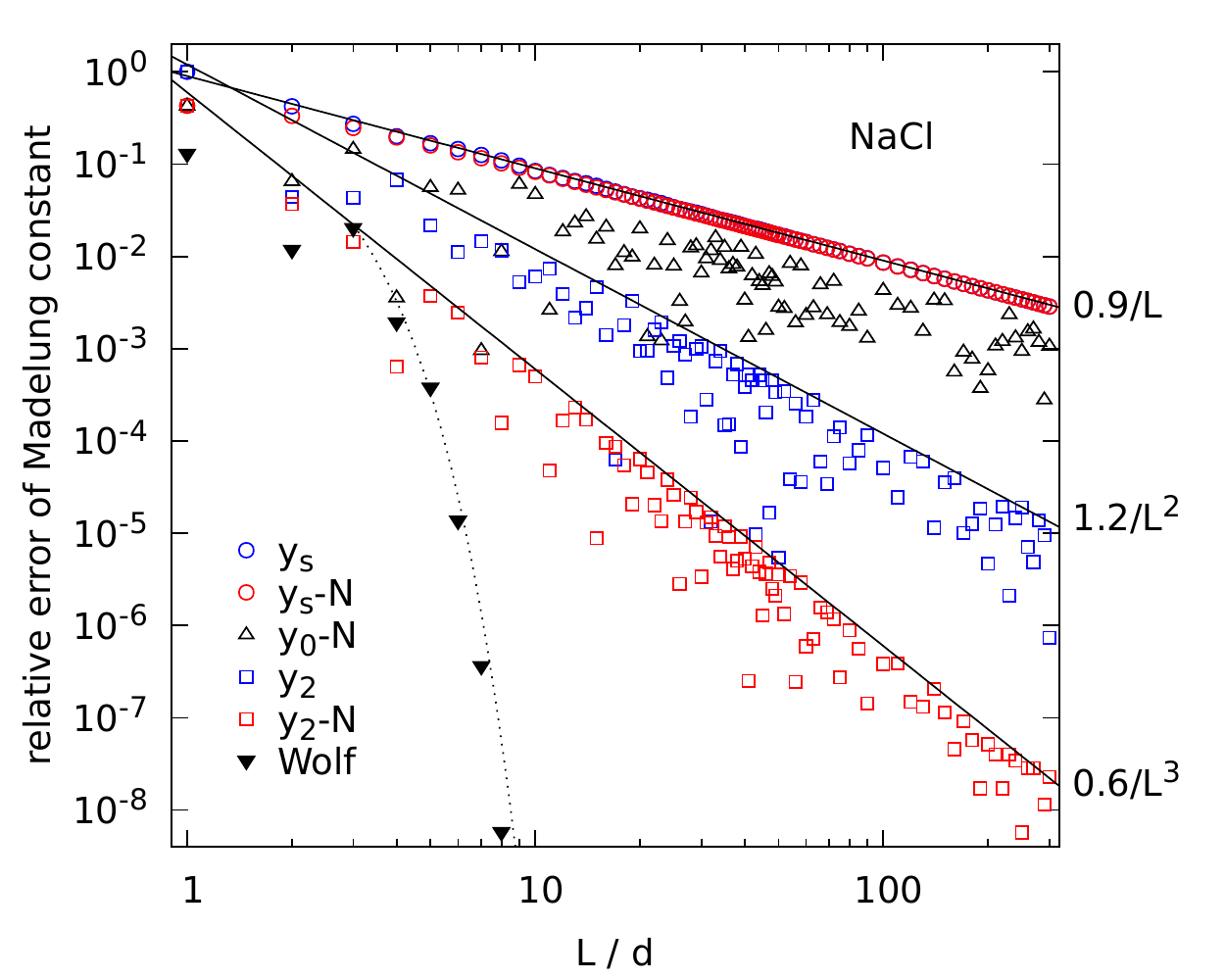}
\caption{Relative error $|M(L)/M^\infty-1|$ of the Madelung constant of NaCl
  computed as function of sphere diameter $L$.
  Finite volume result ($y_s$, blue circles), $y_2$-extrapolation
  (blue squares) and the same with charge neutralization (red
  symbols, $y_s$-N and $y_2$-N), truncated sum with
  charge neutralization ($y_0$-N, up triangles)
  and the Wolf method~\cite{wolf} with damping parameter
  $\alpha=0.5$ (filled down triangles) are compared.
  The solid lines are power law fits of the maximum error
  and the dotted line is a guide to the eye.
}\label{fig:nacl}
\end{center}
\end{figure}

In the following we consider a sphere of diameter~$L$ and calculate,
with $y=y_s$ (Eq.~\ref{ysx}), 
the exact finite volume integrals $H^V_{\alpha\beta}$ (Eq.~\ref{Habv}) 
and the corresponding ensemble averaged Madelung constants
(Eqs~\ref{Mav},\ref{phi3}).
We also compute extrapolations of $H_{\alpha\beta}^V$ to infinite volume,
obtained with $y=y_2$ (Eq.~\ref{y2x}) as well as 
the usual, unweigthed sums truncated at $r=L$,
obtained with $y=y_0$ (Eq.~\ref{y0x}).
The convergence of the Madelung constant of the NaCl structure
is shown in Fig.~\ref{fig:nacl}.
The exact, infinite lattice value is $M^\infty=1.7475646$.
The relative error $|M(L)/M^\infty-1|$ is plotted as a function of
system size~$L$.
The exact finite volume result ($y_s$)
converges very smoothly as $1/L$
while the $y_2$-extrapolation ($y_2$) converges as $1/L^2$.
The truncated sum with $y=y_0$ corresponds to the most simple 
summation order over spherical shells. As it is well known,
this summation order strongly diverges for the NaCl
structure~\cite{borwein85}
and the result is not shown in~Fig.~\ref{fig:nacl}.

We have examined the effect of charge neutralization
using the shifted potential method by Wolf {\it et al.}~\cite{wolf},
which consists in replacing the true Coulomb interaction $q_iq_j/r_{ij}$
by $q_iq_j(1/r_{ij}-1/L)$. For the usual, truncated sum ($y_0$),
this is equivalent to adding a term $-Q/L$ to the potential,
where $Q$ is the total charge in the sphere of
radius~$L$~\cite{wolf,harrison}.
Charge neutralization makes the truncated sum converge, roughly
as as~$1/L$, as seen from the curve $y_0$-N in Fig.~\ref{fig:nacl}.
Interestingly, for the exact finite volume integrals ($y_s$),
charge neutralization has almost no effect. Indeed, the curve
with charge neutralization ($y_s$-N) is almost identical to that without
($y_s$).
This indicates that the positional averaging
(see Fig.~\ref{fig:scheme}), which is implicit in $y_s$,
largely cancels charge imbalances and makes explicit
charge neutralization unnecessary.
For the $y_2$~extrapolation, however, charge neutralization speeds
up convergence considerably and the charge-neutralized sums
($y_2$-N) converge as $1/L^3$, in the same way as $y_2$-extrapolated,
proper KBI integrals~\cite{kruger18}.
For comparison, the Wolf method~\cite{wolf} is also shown
with an Ewald damping parameter $\alpha=0.5$. Because of
the ${\rm erfc}(\alpha r)$-like damping term used in this scheme,
the lattice sum converges exponentially fast
and thus outperforms any method with power-law convergence, such as the
present one. However, the Wolf method requires an empirical
parameter~($\alpha$), whose optimum value depends on the problem at hand.
As seen from Fig.~\ref{fig:nacl}, with the present scheme $y_2$-N,
a practically reasonable error of $10^{-3}$ is
achieved with a very moderate cut-off $L=10d$,
which is comparable to $L=6d$ needed with the Wolf method.
In conclusion of this section we have shown that ensemble averaged
Madelung constants, Eqs~(\ref{Mav},\ref{phi2}),
converge unconditionally even without
charge compensation or damping factors.
The $y_2$-extrapolation with charge neutralization
converges as $1/L^3$, which is quite fast, albeit
slower than the Wolf method.

\section{Size dependence of Madelung energies}
\subsection{Universal surface term}
\begin{figure}[htbp]
\begin{center}
\includegraphics[width=\columnwidth]{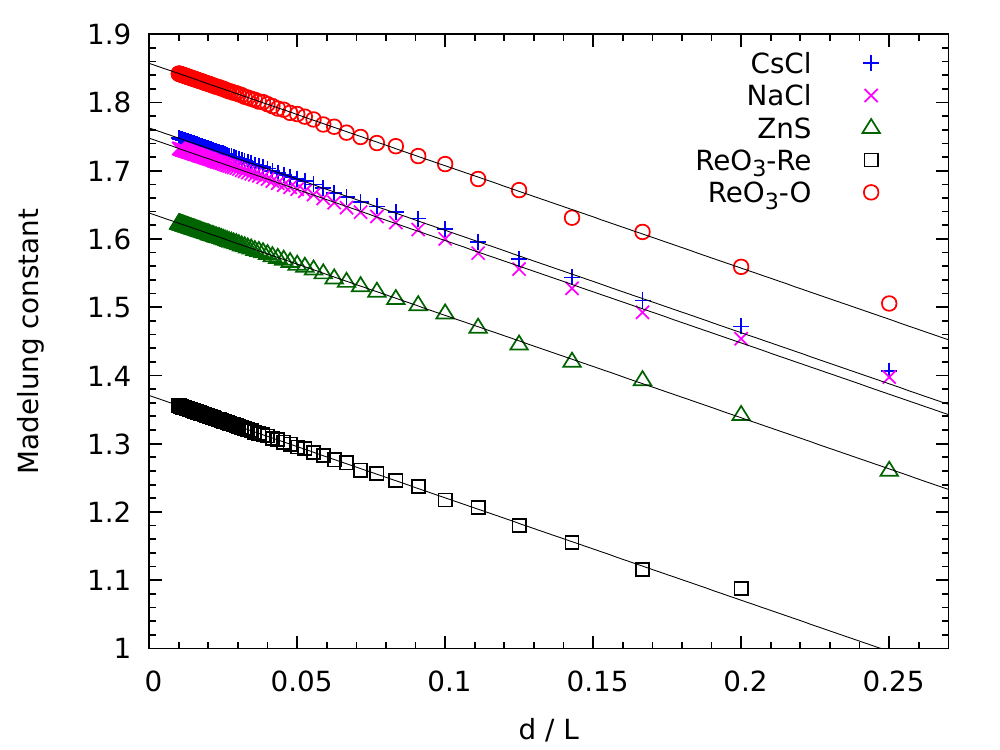}
\caption{
  Ensemble averaged Madelung constant of a sphere of
  diameter $L$, for NaCl, CsCl, ZnS and ReO$_3$ structures
  with nearest neighbor distance~$d$.
  The lines are $M^\infty_\alpha-1.5 d/L$, where $M^\infty_\alpha$
  is the infinite crystal value.
 }\label{fig:univ}
\end{center}
\end{figure}

Finite volume integrals like $H^V$ in Eq.~(\ref{Hw}) can be expanded in
powers of $1/L$ as~\cite{kruger18} 
\begin{equation}\label{hexpand}
  H_{\alpha\beta}(L) = H_{\alpha\beta}^\infty + F_{\alpha\beta}^\infty/L
  + {\cal O}(L^{-2}) \;,
\end{equation}  
where $F^\infty_{\alpha\beta}$ is the surface term in the large volume limit,
given by
\begin{equation}\label{finfty}
F_{\alpha\beta}^\infty = -\frac{3}{2}\int_0^\infty rh_{\alpha\beta}(r)
v(r) 4\pi r^2 dr \;.
\end{equation}
For the Coulomb potential $v(r)=1/r$, this simplifies to
$F_{\alpha\beta}^\infty = -3G_{\alpha\beta}^\infty/2$, where
\begin{equation}\label{ginfty}
  G_{\alpha\beta}^\infty=\int_0^\infty h_{\alpha\beta}(r)
4\pi r^2 dr 
\end{equation}
is a proper KBI. The latter is directly related to the
particle number fluctuations in the volume~$V$ as~\cite{bennaim}
\begin{equation}\label{rhoG}
  \rho_\alpha G_{\alpha\beta}^\infty=
  \frac{\langle N_\alpha N_\beta\rangle
    -\langle N_\alpha\rangle\langle N_\beta\rangle}{\langle N_\beta\rangle}
    -\delta_{\alpha\beta} \;.
\end{equation}
Here we are interested in a solid at low temperature
where the fluctuations are negligible, i.e.
$\rho_\alpha G_{\alpha\beta}^\infty= -\delta_{\alpha\beta}$
and so
\begin{equation}\label{fdelta}
  F_{\alpha\beta}^\infty = ({3}/{2})\delta_{\alpha\beta}/\rho_\alpha \;.
\end{equation}
Using Eqs.~(\ref{phi3},\ref{hexpand},\ref{fdelta}),
the size dependence of the ensemble averaged
Madelung constants~(\ref{Mav}) is found to be
\begin{equation}\label{ML}
  M_\alpha(L) = M_\alpha^\infty -({3}/{2})d/L + {\cal O}(L^{-2}) \;,
\end{equation}
which holds for all ionic species and {\em for any structure}.
This is exemplified with a few crystal structures types in Fig.~\ref{fig:univ}.
It can be seen that all Madelung constants approach the infinite volume
limit with the same slope $-3/2$,
confirming the general validity of Eq.~(\ref{ML}).

From Eqs~(\ref{U},\ref{ML}) and $L=6V/A$,
the surface contribution to the Madelung energy, per area~$A$, 
is given by
\begin{equation}\label{us}
  U^S  = \rho \langle{q^2}\rangle /8 \;,
\end{equation}
where $\rho=\sum_\alpha\rho_\alpha$ is the total particle density and
\begin{equation}\label{q2}
  \langle{q^2}\rangle=\sum_\alpha N_\alpha q_\alpha^2 /N
\end{equation}
is the mean square charge of the ions.
We stress that Eq.~(\ref{us}) corresponds to the spherical average over
all surface orientations.
As we have made no assumption about the crystal structure,
this result is universal, i.e.\
it holds for any crystal and amorphous structure.
For binary systems, it simplifies to $U^S= \rho q^2/8$.

The ensemble averaged surface energy of Eq.~(\ref{us})
can be considerably larger than the surface energies found
in real ionic crystals.
For NaCl, for example, Eq.~(\ref{us}) gives 1.28~J/m$^2$,
which is several times larger than the experimental values
reported for the low index surfaces (100) and (110),
which are 0.15--0.18 J/m$^2$ and 0.35--0.45 J/m$^2$,
respectively~\cite{tasker}.
The main reason for this discrepancy is that $U^S$ is an
average over all surface orientations, which includes
all kinds of vicinal surfaces as well as highly unstable polar faces.
Real crystals, however, are faceted,
and only the most stable, low index surfaces are actually present
in crystallites.
Second, ionic and electronic relaxation and reconstruction,
not considered here, make the surface energy decrease further,
especially for stepped and polar terminations.

\subsection{Relative stability of finite clusters}
Despite the fact that Eq.~(\ref{us}) overestimates
the surface energy of ionic crystals, the present findings,
which are valid for all ionic systems,
provide important insight into the size dependence
of the Madelung energy. 
In particular, we can understand trends in the relative stability
of different structures as a function of system size.
From Eqs.~(\ref{U},\ref{ML}) the ensemble averaged Madelung energy of
a subvolume~$V$ with an average of $N=\rho V$ ions is given,
to first order in $1/L$, by
\begin{equation}\label{UL}
  U = -\frac{N}{2d}\sum_\alpha c_\alpha q_\alpha^2
  \left( M_\alpha^\infty - \frac{3d}{2L} \right) \;,
\end{equation}
where $c_\alpha=N_\alpha/N$ is the concentration of
$\alpha$~ions.
Using Eq.~(\ref{UL}) the ensemble averaged Madelung energy of finite
clusters of different crystal structures can be compared. 
We write $d/L = \chi/N^{1/3}$,
where $\chi$ is a geometrical factor that only depends on the
crystal structure and the shape of the subvolume.
In the following we consider
binary structures, where $c_+=c_-=1/2$, $q_+=-q_-$ and $M_+=M_-$.
As the most simple example, we compare two ionic structures with
the same nearest neighbor distance~$d$, which is a reasonable assumption
when cations and anions have about the same ionic radius.
From Eq.~(\ref{UL}) we then find the energy difference between the two
clusters
\begin{equation}\label{DU}
  \Delta U = -\frac{Nq^2}{2d}\left(
  \Delta M^\infty - \frac{3\Delta\chi}{2N^{1/3}} \right) \;.
\end{equation}
where $\Delta X = X_1-X_2$  ($X=U,M^\infty,\chi$).
We label the two structures such that $\Delta M^\infty>0$,
i.e. structure~1 is the stable phase in the bulk.
If $\Delta\chi>0$, then $\Delta U$ changes sign at the cluster size
\begin{equation}\label{N0}
  N_0=\left(\frac{3\Delta\chi}{2\Delta M^\infty}\right)^3\;.
\end{equation}
Then, structure~1 is stable for $N>N_0$ and
structure~2 for $N<N_0$.
As an example, we compare CsCl (structure~1) and NaCl (structure~2).
We have $M^\infty_1 = 1.762675$ (CsCl), $M^\infty_2=1.747565$ (NaCl)
and so $\Delta M^\infty=0.015110$.
We consider spherical volumes.
It is easy to see from simple geometrical relations that
$\chi_1=\sqrt{3}/2\times(\pi/3)^{1/3}$ (CsCl)
and $\chi_2=(\pi/6)^{1/3}$ (NaCl),
which gives $\Delta\chi=0.073445$.
From Eq.~(\ref{N0}) we then obtain~$N_0=388$.
So the model predicts a phase transition from the NaCl structure
to the CsCl structure when the cluster size exceeds about 400 atoms.
The model may be applied to CsBr, CsCl and CsI,
where the ionic radius of cation and anion is about the same
(Cs=1.81\AA, Cl=1.67\AA, Br=1.82\AA, I=2.06\AA).
In the bulk, these compounds crystallize in the CsCl structure.
For small clusters, however, the NaCl structure has been observed 
in experiment~\cite{twu90,hautala17} and found to be
more stable in first principles calculations~\cite{aguado98,aguado00}.
According to a recent experiment on neutral CsBr clusters,
the transition from the NaCl to the CsCl structure
occurs for a cluster size of about 160 atoms~\cite{hautala17},
in good agreement with our simple model.

\section{Conclusions}
In summary, we have developed a theory of ensemble
averaged Madelung energies in finite volumes.
The Madelung constants approach the thermodynamic limit
in a universal way, independent of the structure, may it be
crystalline or amorphous.
A simple, general expression has been derived for
the Madelung part of the termination averaged surface energy.
The size dependence of the Madelung energies
helps to understand the relative stability of different ionic
structures as a function of system size, and provides
a simple explanation for the phase transition from the rocksalt
to the cesium chloride structure, which is observed in CsX clusters
(X=Cl,Br,I).

\subsection*{Acknowledgments}
I thank Jean-Marc Simon and Thijs Vlugt for stimulating
discussions. This work was supported by JSPS KAKENHI Grant Number 19K05383.

\end{document}